\def\figcond{1}     
\def\figs#1#2#3#4#5#6#7{%
\begin{figure}[#2]
\ifnum\figcond>0
  \centerline{\hbox{\hsize=6in\hss\vbox to #5in{\vss%
  \vskip #6in\centerline{\hskip #4in\epsfig{file=#1,width=#3in}}
  \vss}\hss}}
\else
  \centerline{\hbox{\hsize=6in\vrule\hss\vbox to #5in{\hrule\vss%
  \centerline{Figures or Hardcopys available from %
              koepf@albert.tau.ac.il}%
  \vss\hrule}\hss\vrule}}
\fi
\vskip 0.15in
\caption{#7}
\end{figure}}
\ifnum\figcond>0
\documentstyle[preprint,floats,prc,aps,epsfig]{revtex}
\else
\documentstyle[preprint,floats,prc,aps]{revtex}
\fi
\begin{document}
\preprint{TAUP-2273-95, hep-ph/9507218}
\draft

\title{THE NUCLEON'S VIRTUAL MESON CLOUD AND DEEP INELASTIC LEPTON
       SCATTERING}

\author{W. Koepf and L.L. Frankfurt\thanks{On leave of absence from
the St.~Petersburg Nuclear Physics Institute, Russia.}}
\address{School of Physics and Astronomy\\
         Raymond and Beverly Sackler Faculty of Exact Sciences\\
         Tel Aviv University, 69978 Ramat Aviv, Israel}

\author{M. Strikman\thanks{Also St.~Petersburg Nuclear Physics
Institute, Russia.}}
\address{Pennsylvania State University, University Park, PA 16802,
USA}

\date{July 4, 1995}

\maketitle
\begin{abstract}

We address the question whether the nucleon's antiquark sea can be attributed
entirely to its virtual meson cloud and, in essence, whether there exists a
smooth transition between hadronic and quark-gluon degrees of freedom. We
take into account contributions from $\pi$ and $K$ mesons and compare with
the nucleon's antiquark distributions which serve as a non-perturbative
input to the QCD evolution equations. We elucidate the different behavior in
the flavor singlet and non-singlet channels and study the dependence of our
results on the scale $Q^2$. The meson-nucleon cut-offs that we determine give
not only an indication on the size of the region within which quarks are
confined in a nucleon, but we find that the scale of these form factors is
closely related to the four-momentum transfer, $Q^2$, where gluons are
resolved by a high energy probe, and that large meson loop momenta, $|{\bf
k}| \approx 0.8$ GeV, contribute significantly to the sea quark distributions.
While the agreement of our calculations with data-based parametrizations is
satisfactory and scale independent for the flavor breaking share of the
nucleon's antiquark sea, the flavor singlet component is quite poorly
described. This hints the importance of gluon degrees of freedom.

\end{abstract}
\bigskip
\section{INTRODUCTION}

Ever since the postulation of mesons by Yukawa in 1934, and the
discovery of the pion in 1947, it has been clear that mesons play a
crucial role in low-energy nuclear physics. The long-range part of
the nucleon-nucleon interaction at low energies is clearly dominated
by one-pion exchange, and direct evidence for the role of mesons in
nucleon structure comes from the negative charge radius of the
neutron which can be attributed to the virtual $n\to p\pi^-$ process.
Furthermore, meson-exchange currents have proven to be essential for
a quantitative description of many low-momentum-transfer processes,
as, for instance, radiative neutron capture at threshold,
near-threshold electro-disintegration of the deuteron, and the
isovector magnetic form factors of nuclei. Also, a non-perturbative
pionic cloud around the nucleon offers a straightforward explanation
\cite{henl90} of the $SU(2)$ flavor asymmetry of the proton's
antiquark sea observed in the NMC experiment \cite{amau91}.

The fundamental role of pion clouds in nuclear physics is well
explained in QCD as a consequence of the spontaneously broken chiral
symmetry, and an interesting and important question is on the region
of applicability of Chiral QCD Lagrangians, and at what distance
scale this description will fail. It is the purpose of this work to
investigate whether there exists a smooth transition between
low-energy nuclear physics degrees of freedom, i.e., baryons and
mesons, and  a description in terms of quarks and gluons which is
adequate for hard processes. In particular, we study whether the
nucleon's antiquark distributions, as observed in inclusive,
unpolarized, deep inelastic lepton and neutrino scattering, can be
attributed entirely to its virtual meson cloud or if they should be
described in terms of quarks and gluons. For this aim, we compare the
contribution of the meson cloud with primary sea quark distributions
which serve as a non-perturbative input to the QCD evolution
equations.

In traditional nuclear physics, the crucial quantity which determines
the strength of the pionic contribution to the nucleon's structure is
the $\pi N N$ form factor. In a quark model picture, the latter is
intimately related to the confinement size of the quarks, and there
is a long standing theoretical puzzle associated with it. The need
for a sufficiently strong tensor force to reproduce the $D/S$ ratio
and the quadrupole moment of the deuteron suggests a rather hard $\pi
NN$ vertex at small momentum transfers,
and consequently most $NN$ meson-exchange potentials
which are fit to the rich body of $NN$ phase shift data have a
relatively high momentum-cutoff, with, for example, $\Lambda_{\pi
NN}=1.3$ GeV in monopole form for the Bonn potential \cite{mach87}.

On the contrary, hadronic models of baryons with meson clouds, like
the Skyrmion, typically have a rather soft $\pi NN$ form factor
\cite{bali87}, in a quenched lattice QCD calculation a soft form
factor was "measured" with a monopole mass of $\Lambda_{\pi NN}=0.75
\pm 0.14$ GeV \cite{kliu94}, and also in a recent analysis in the
framework of QCD sum rules a soft monopole cut-off of $\Lambda_{\pi
NN} \approx 0.8$ GeV was suggested \cite{meis95}. In addition,
there is further evidence for a fairly soft $\pi NN$ vertex from
other sources: arguments based on resolving the Goldberger-Treiman
discrepancy \cite{coon81} as well as the apparent charge dependence
of the $pp\pi^0$ and $pn\pi^+$ couplings \cite{thom89} both suggest a
relatively soft $\pi NN$ form factor with $\Lambda_{\pi NN} \approx
0.8$ GeV. Threshold pion production from $pp$ scattering can best be
reproduced by using a soft $\pi NN$ vertex, with $\Lambda_{\pi NN}
\approx 0.65$ GeV \cite{hlee95}, and pion electro-production data on
hydrogen also point towards a very soft $\pi NN$ form factor
\cite{guet84}.
Today, there exist efforts to reconcile the $NN$ phase shift data
with a soft $\pi NN$ vertex. The inclusion of $\pi\pi$ interactions
\cite{hkim94} as well as $\pi\rho$ scattering \cite{jans93} allows
one to avoid the need for hard form factors,
and Haidenbauer et al.~\cite{haid94}
presented a successful model where both one- and two-pion exchanges
were included with soft $\pi NN$ and $\pi N\Delta$ vertices.

Furthermore, a soft $\Lambda_{\pi NN}$ is preferable to avoid
contradictions with data on the nucleon's antiquark distributions.
Thomas \cite{thom83} pointed out that deep inelastic
scattering data on integrals over the momentum carried by sea quarks
in the proton, i.e., sum rules, can be used to restrict the
$t$-dependence of the $\pi NN$ vertex. Frankfurt et al.~\cite{fran89}
showed that to describe the steep decrease of the sea quark
distributions with $x$ the vertex cut-off, in a monopole
parametrization,
should be less than 0.5 GeV. Subsequently, sparked by experimental
results of the NMC group \cite{amau91} which suggested a violation
of the Gottfried sum rule the analysis of this mechanism was focused
on the $SU(2)$ breaking component of the quark sea
\cite{henl90,kuma91,sign91,meln91,zoll92,shak94}, and more mesons
were included into the nucleon's virtual cloud
\cite{hwan91,szcz93,meln93}.

In many early works in that realm solely integrated quantities, i.e.,
sum rules, were discussed \cite{henl90,thom83,sign91}, and in others
either only the flavor breaking share of the nucleon's antiquark
sea was considered \cite{kuma91,meln91,zoll92,shak94} or the
analysis was limited to certain combinations of the nucleon's
sea quark distributions \cite{fran89,meln93}. In
Refs.~\cite{hwan91,szcz93}, the nucleon's entire antiquark sea was
attributed to its meson cloud at a scale of $Q^2 \approx 17$ GeV$^2$
while using hard vertices which are almost consistent with
the Bonn meson-exchange model. Their conjecture, however, is based
predominantly on integrated quantities (sum rules), and the
contribution of the mesonic cloud was multiplied with a wave function
renormalization factor which is at variance with the standard
nuclear physics definition of the $\pi NN$ coupling constant
\cite{meln93}, as will be clarified later. Besides, to describe the
flavor asymmetry in the sea quark distributions specific assumptions
on the quark distributions in the bare, recoiling baryons were needed.

Also in this work, we investigate the possibility to attribute the
nucleon's full antiquark sea to its virtual mesonic cloud.
However, we do not consider sum rules which contain
contributions from the small $x$ region where shadowing effects are
important, but we study the $x$-dependence of the antiquark
distributions, emphasizing especially the tails of the parton
distribution functions which are most sensitive to the meson-nucleon
form factors.

For the first time, the dependence of the parameters which
characterize the underlying convolution picture, i.e., the cut-offs
at the vertices, on the scale, $Q^2$, where perturbative QCD evolution
of the non-perturbative sea quark distributions starts is studied,
and the qualitatively different behavior in the flavor singlet and
non-singlet channels is elucidated. Also, we analyze which mesonic
virtualities and loop momenta yield the dominant contributions in
the respective convolution integrals, and, in essence, whether there
exists a smooth transition between hadronic and quark-gluon degrees
of freedom.

Recently, new improved fits of the nucleon's unpolarized parton
distributions to a host of deep inelastic scattering data became
available from Martin, Roberts and Stirling \cite{mart95} as well as
the CTEQ collaboration \cite{cteq94}. Using those, we update the
analysis of the contribution of the virtual meson cloud of the
nucleon to its antiquark distributions, and separately adjust the
$\pi NN$, $KNY$ (where $Y \in \{\Lambda, \Sigma, \Sigma^*\}$) and
$\pi N\Delta$ form factors to the flavor $SU(3)$ and $SU(2)$ breaking
components of the nucleon's antiquark sea and to its strange quark
content, following the framework presented, for instance, in
Refs.~\cite{henl90,thom83,fran89,sull72}.

In particular, we fit the various meson-nucleon vertices to the
nucleon's antiquark distributions at different values of the
four-momentum transfer, $Q^2$, where the contamination from gluon
splitting into $q\bar q$ pairs is not yet dominant. In addition to
the major contribution arising from virtual pions, we also consider
the kaonic cloud, with the corresponding coupling constants fixed by
spin-flavor $SU(6)$ which holds much better for the coupling
constants than it does for the masses, as has been observed in
hyperon-nucleon scattering \cite{holz89}. However, other mesons whose
contributions are even more suppressed as due to their higher masses
were not considered.

The form factors that we determine in this work are still
significantly softer than what is used in most meson-exchange models
of the $NN$ interaction, and they indicate that the $\pi N\Delta$
vertex is considerably softer than the $\pi NN$ vertex and that the
cut-offs in the strange sector are harder than in the non-strange
sectors. We analyze the relationship between the
hardness of the meson-nucleon form factors and $Q^2$, the scale where
QCD evolution begins. While the agreement of our calculations with
the data-based parametrizations is satisfactory and practically scale
independent for the flavor breaking share of the nucleon's
antiquark distributions, the corresponding flavor singlet component
is quite poorly described in the convolution picture and it is
contaminated with a sound scale dependence. The meson virtualities
and loop momenta that are relevant in the deep inelastic process
investigated here, $t \approx -0.4$ GeV$^2$ and $|{\bf k}| \approx
0.8$ GeV, are very different from those probed in low-energy nuclear
physics phenomena, as, for instance, in the meson-exchange
descriptions of the $NN$ interaction where meson momenta of the order
of the pion mass are dominant, and the validity of Chiral QCD
Lagrangians in this regime is questionable. Actually, the analysis
performed here hints that hard form factors at the
meson-nucleon vertex may reflect the presence of a gluon cloud in the
nucleon which cannot be resolved by a low energy probe, and that
there exists no smooth transition between hadronic and quark-gluon
degrees of freedom.

The organization of this work is as follows. In Sec.~II, we review
the theoretical framework with which we attempt to connect the
nucleon's sea quark distributions to its virtual meson cloud. In
Sec.~III, we present results of our fits to the flavor $SU(3)$
and $SU(2)$ breaking components of the nucleon's antiquark
sea and to its strange quark content.  We discuss the results of our
calculations in Sec.~IV, and summarize and conclude in Sec.~V.

\section{THE PION CLOUD AND THE NUCLEON'S STRUCTURE FUNCTIONS}

It has been suggested by Sullivan \cite{sull72} that in lepton
scattering the virtual one-pion exchange may give a significant
contribution to the nucleon's deep inelastic structure functions
which scales in the Bjorken limit like the original process, and the
generalization of that mechanism is depicted in Fig.~1.

\figs{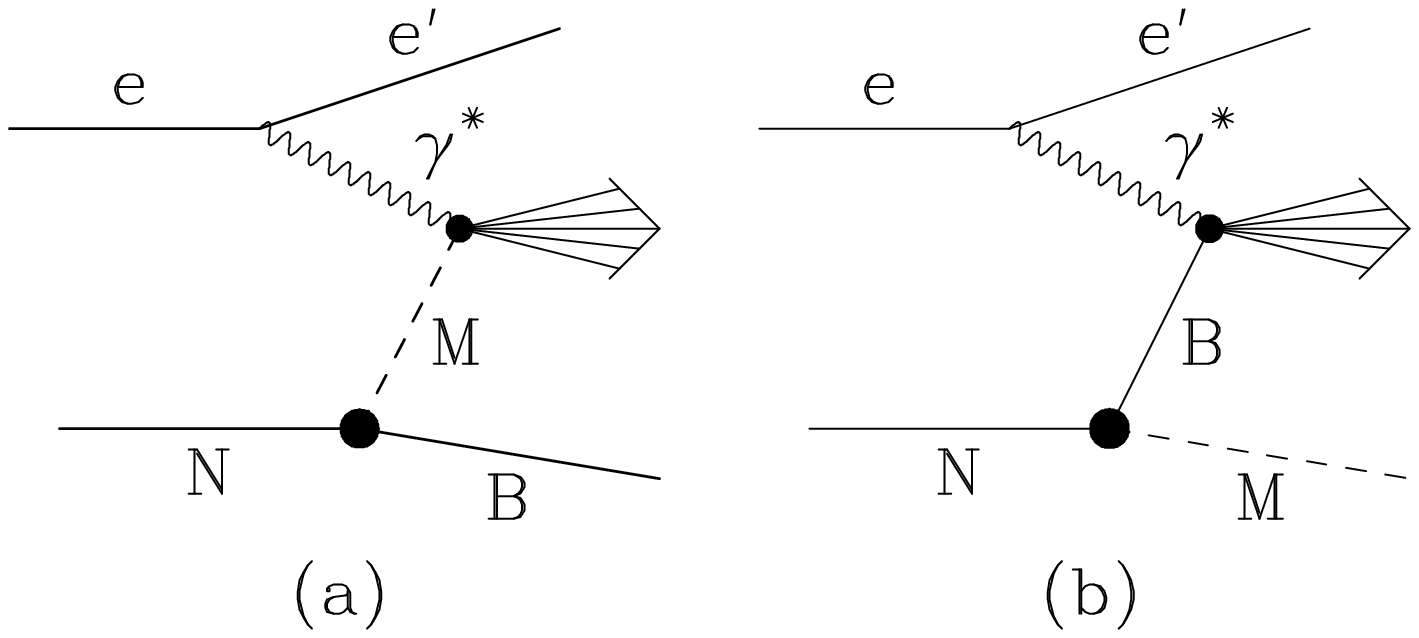}{htbp}{6.0}{0.0}{2.1}{0.8}{The meson cloud contribution
to the nucleon's structure functions in deep inelastic lepton
scattering, where $B \in \{N, \Delta, \Lambda, \Sigma, \Sigma^*\}$
refers to an octet or a decuplet baryon accessible from the nucleon
through emission of a meson, $M \in \{\pi , K\}$.}

However, it is also common wisdom that, at large $Q^2$, a significant
fraction of the nucleon's antiquark sea originates from gluon
splitting into $q \bar q$ pairs and not from the mesonic cloud. This
perturbative process is approximately flavor symmetric and hence
does not contribute to the flavor $SU(3)$,
\begin{equation}
x \bar q_8(x,Q^2) \equiv
      x \left[ \bar u(x,Q^2) + \bar d(x,Q^2) - 2 \bar s(x,Q^2) \right]
\ ,
\label{eq:su3}
\end{equation}
and $SU(2)$,
\begin{equation}
x \bar q_3(x,Q^2) \equiv
      x \left[ \bar d(x,Q^2) - \bar u(x,Q^2) \right]
\ ,
\label{eq:su2}
\end{equation}
breaking components of the nucleon's antiquark distributions.

The direct meson cloud contribution of Fig.~1a can be written as a
convolution of the virtual meson's antiquark distribution and its
momentum dependence in the infinite momentum frame,
\begin{equation}
x \, \bar q_N(x,Q^2) = \sum_{M,B} \alpha_{MB}^q
\int^1_x dy \, f_{MB} (y) \,
{x \over y} \, \bar q_M\!\left({x \over y},Q^2\right)
\ ,
\label{eq:con}
\end{equation}
where, in our investigation, the sum runs over $M \in \{\pi ,K\}$ and
$B \in \{N,\Delta ,\Lambda ,\Sigma ,\Sigma^*\}$, and where the
$\alpha_{MB}^q$ are spin-flavor $SU(6)$ Clebsch-Gordan factors. We
assume the quark sea in the mesons to be flavor symmetric, and thus
only the $\pi$ and K {\em valence antiquark} distributions contribute
to Eqs.~(\ref{eq:su3}) and (\ref{eq:su2}). We neglect the slight
difference between the latter, and take $x \bar q_\pi (x,Q^2)$ from
fits to Drell-Yan pair production experiments. We also disregard
contributions from mesons other than the $\pi$ and the K which are
strongly suppressed due to their higher masses as well as
interference effects, as, for instance, between the $\pi$ and the
$\rho$. Actually, the latter vanish for the unpolarized
distributions, to which we limit our discussions, because the
corresponding trace over the baryon spinors and a pseudoscalar and a
vector vertex is identically zero.

The virtual meson's light-cone distribution in the nucleon's cloud,
\begin{equation}
f_M (y) = \sum_B f_{MB} (y)
\ ,
\label{eq:fmy}
\end{equation}
which characterizes its probability of carrying a fraction
$y$ of the nucleon's momentum in the infinite momentum frame,
can be be expressed as
\begin{equation}
f_{MB}(y) = {g_{MNB}^2 \over 16\pi^2} \, y \, \int_{-\infty}^{t_{min}}
             dt \, {{\cal I}(t,m_N,m_B) \over (t-m_M^2)^2} \,
             F^2_{MNB}(t)
\ ,
\label{eq:fmb}
\end{equation}
where
\begin{equation}
{\cal I}(t,m_N,m_B) = \left\{
\begin{array}{ll}
    -t+(m_B-m_N)^2
    & \qquad \mbox{for $B \in$ {\bf 8}} \\
    \noalign{\medskip}
    {\big((m_B+m_N)^2-t\big)^2 \big((m_B-m_N)^2-t\big)
    \over 12 m_N^2 m_B^2 }
    & \qquad \mbox{for $B \in$ {\bf 10}} \ ,
\end{array} \right.
\label{eq:itb}
\end{equation}
for an intermediate, on-mass-shell octet or decuplet baryon,
respectively, and where for the latter a Rarita-Schwinger spin
vector was employed when evaluating the trace over the pseudoscalar
vertices.

The integration in Eq.~(\ref{eq:fmb}) is over the meson's virtuality,
$t=k^2$, where $k=(m_N-\sqrt{m_B^2+|{\bf k}|^2},{\bf k})$ is its
four-momentum, and the upper limit of the integration is determined
purely by kinematics, with
\begin{equation}
t_{min} = m_N^2 y - {m_B^2 y \over 1-y}
\ .
\label{eq:tmx}
\end{equation}
Here, $m_N$, $m_B$ and $m_M$ are the nucleon mass, the mass of the
intermediate baryon and the meson mass. For the pion-nucleon
couplings we use the most recent values of the pseudoscalar coupling
constants of ${g_{\pi NN}\over\sqrt{3}}=13.05$ \cite{stok93} and
$g_{\pi N\Delta}=28.6$ \cite{davi90}, and the kaon-nucleon couplings
are related to the latter using spin-flavor $SU(6)$ which agrees well
with hyperon-nucleon scattering \cite{holz89}.

The only, a priori, unknown quantity in Eq.~(\ref{eq:fmb}) is then
the form factor, $F_{MNB}(t)$, which governs the emission of an
off-mass-shell meson, and it is usually parametrized either in
monopole, dipole or exponential (Gaussian) form, where
\begin{equation}
F_{MNB}(t) = \left\{
\begin{array}{ll}
    {\Lambda_m^2-m_M^2 \over \Lambda_m^2-t}
    & \qquad \mbox{monopole} \\
    \noalign{\medskip}
    \left({\Lambda_d^2-m_M^2 \over \Lambda_d^2-t}\right)^{\!\!2}
    & \qquad \mbox{dipole} \\
    \noalign{\medskip}
    e^{(t-m_M^2)/\Lambda_e^2}
    & \qquad \mbox{exponential} \ ,
\end{array} \right.
\label{eq:for}
\end{equation}
and where the cut-off masses can depend, in general, also on the
meson-baryon channel under consideration. For our purposes,
differences between the various forms given in Eq.~(\ref{eq:for}) are
not particularly important, and we will translate between the
different cut-off parameters using the approximate relation given by
Kumano \cite{kuma91},
\begin{equation}
\Lambda_m \approx 0.62 \Lambda_d \approx 0.78 \Lambda_e
\ ,
\label{eq:kum}
\end{equation}
which is based on demanding an identical reduction of the various
form factors of Eq.~(\ref{eq:for}) to 40\% of their pole values,
i.e., $F^m_{MNB}(t_0)=F^d_{MNB}(t_0)=F^e_{MNB}(t_0)=0.4$. With this,
the form factors are compared at large virtualities, $t_0 \approx -
\Lambda^2$. This is in contrast to the standard procedure of relating
the different cut-off parameters by means of the slopes of the form
factors at the meson poles, which, conversely, would lead to
\begin{equation}
\Lambda_m \approx {\Lambda_d \over \sqrt{2}} \approx \Lambda_e
\ ,
\label{eq:der}
\end{equation}
and which is a good approximation for small virtualities. As the
major contributions to the convolution integrals stem from fairly
large mesonic virtualities of $t \approx -0.4$ GeV$^2$, as will be
discussed in Sect.~IV.B, Eq.~(\ref{eq:kum}) is more appropriate
in this realm than Eq.~(\ref{eq:der}). Because for a monopole form
factor the contribution to Eq.~(\ref{eq:fmb}) from intermediate
decuplet baryons is UV-divergent \cite{fran89}, we restrict ourselves
to an exponential form in our actual calculations, which, in
addition, yields the most satisfactory results.

We finally obtain for the flavor breaking components of the nucleon's
antiquark distributions,
\begin{mathletters}
\begin{eqnarray}
x \, \bar q_8(x,Q^2) &=& \int^1_x dy \,
\left[ \, f_{\pi N} (y) + f_{\pi \Delta} (y) - 2 f_K (y) \, \right]
\, {x \over y} \, \bar q_\pi\!\left({x \over y},Q^2\right) \ ,
\label{eq:su3a}
\\
\noalign{\medskip}
x \, \bar q_3(x,Q^2) &=& \int^1_x dy \,
\left[ \, {2 \over 3} f_{\pi N} (y) -
       {1 \over 3} f_{\pi \Delta} (y) \, \right]
\, {x \over y} \, \bar q_\pi\!\left({x \over y},Q^2\right) \ ,
\label{eq:su2a}
\end{eqnarray}
\end{mathletters}
with $f_{\pi N}(y)$ and $f_{\pi \Delta}(y)$ from Eq.~(\ref{eq:fmb}),
and $f_K(y)$ from Eq.~(\ref{eq:fmy}). The fact, that the $N \to N\pi$
and $N \to \Delta\pi$ contributions add for the octet component in
Eq.~(\ref{eq:su3a}) while they partially cancel for the triplet
component in Eq.~(\ref{eq:su2a}) offers the possibility to separately
determine $\Lambda_{\pi NN}$ and $\Lambda_{\pi N\Delta}$, as will be
shown in the next section.

The theoretical framework presented in the above agrees with that
employed in Refs.~\cite{henl90,sull72,thom83,fran89,kuma91,sign91},
except that we also consider kaonic loops and that we vary the form
factors in the various meson-baryon channels independently.
In Refs.~\cite{meln91,zoll92,hwan91,szcz93,buch94}, on the other
hand, the Sullivan contribution, depicted in Fig.~1, was in addition
multiplied with a wave function renormalization factor, $Z<1$, where
\begin{equation}
Z = \left( 1 + \sum_M n_M \right)^{\!\!-1}
  = \left( 1 + \sum_M \int_0^1 dy f_M(y) \right)^{\!\!-1}
\ .
\label{eq:zz1}
\end{equation}
This prescription is at variance with the fact that in coordinate
space at large distances the meson-loop diagrams in Fig.~1 describe
physical processes whose cross-sections involve physical,
renormalized couplings. Note that this observation is used
in nuclear physics to fix the $\pi NN$ coupling constant.

In the underlying two-phase model, the physical nucleon, $|N\rangle$,
is pictured as being part of the time a bare core, $|N_0\rangle$, and
part of the time a baryon with one meson "in the air", $|BM\rangle$.
In quantum field theory, bare, unrenormalized couplings, $g_{MNB}^0$,
should be used in the wave function of a physical particle when
expressed in terms of its constituents, i.e.,
\begin{equation}
|N\rangle =
\sqrt{Z'} \left( |N_0\rangle + \sum_M g_{MNB}^0 |BM\rangle \right)
\ .
\label{eq:nn2}
\end{equation}
Bare and renormalized couplings are, to lowest order, related via
$g_{MNB} = \sqrt{Z'} g_{MNB}^0$, which allows us to
rewrite Eq.~(\ref{eq:nn2}) as \cite{theb82}
\begin{equation}
|N\rangle =
\sqrt{Z'} \, |N_0\rangle + \sum_M g_{MNB} |BM\rangle
\ ,
\label{eq:nn3}
\end{equation}
where the wave function renormalization factor,
\begin{equation}
Z' = 1 - \sum_M n_M
   = 1 - \sum_M \int_0^1 dy f_M(y)
\ ,
\label{eq:zz2}
\end{equation}
now only affects the bare core, $|N_0\rangle$, and no longer the
Sullivan contribution, $|BM\rangle$, as discussed in detail in
Ref.~\cite{meln93}. This is the prescription which
we follow, and it is consistent with the standard nuclear physics
definition of the $\pi NN$ coupling constant derived from the $NN$
interaction at large distances.

Instead of the covariant formalism outlined here, in
Refs.~\cite{zoll92} and \cite{meln93} "old-fashioned"
time-ordered perturbation
theory in the infinite momentum frame was used. The two diagrams in
Fig.~1 are then treated on an equal footing, as the active particle
and the spectator are both on their respective mass-shells. However,
energy is not conserved at the meson-nucleon vertices, and the form
factors, $F_{MNB}$, are unknown. In the covariant formalism, it is
thus more straightforward to compare the vertices -- in the form of
$F_{MNB}(t)$ of Eq.~(\ref{eq:for}) -- to standard nuclear physics
quantities as, for instance, the vertex cut-offs of the Bonn
potential.

\section{THE NUCLEON'S ANTIQUARK SEA}

\subsection{The flavor non-singlet components:}

After having outlined the theoretical framework, we present and
discuss the results of our numerical calculations. In the first part
of this section, we limit ourselves to the flavor breaking
components, $x \bar q_8(x,Q^2)$ and $x \bar q_3(x,Q^2)$, defined in
Eqs.~(\ref{eq:su3}) and (\ref{eq:su2}). We determine the
meson-nucleon cut-offs through the $x$-dependence of the various
flavor non-singlet antiquark distributions in the nucleon
by means of the convolution picture outlined in the above, and we
compare results obtained at different scales, $Q^2$, of 1, 2 and
4 GeV$^2$. For the flavor non-singlet components $x \bar q_8(x,Q^2)$
and $x \bar q_3(x,Q^2)$ there is practically no contribution from
gluon splitting into $q \bar q$ pairs, which is approximately flavor
symmetric. Therefore, in this channel, the meson cloud picture has a
good chance to describe the respective antiquark distributions, at
least at a small scale of, for instance, $Q^2=1$ GeV$^2$.

In Fig.~2, we show a comparison of our meson cloud calculations
with various recent, empirical parton distribution functions, MRS(A')
\cite{mart95}, CTEQ3M \cite{cteq94}, GRV94(HO) \cite{glrv94},
BM(A) \cite{berg93} and MRS(D'$_-$) \cite{mart93}, fit to the host of
inclusive, unpolarized, deep inelastic lepton and neutrino scattering
data.  The mesonic contributions, as given in Eqs.~(\ref{eq:su3a})
and (\ref{eq:su2a}), were obtained by evaluating the mesons'
light-cone distributions in the nucleon's cloud with an exponential
form factor using cut-off masses, $\Lambda_e = \Lambda^e_{\pi NN} =
\Lambda^e_{KNY} = \Lambda^e_{\pi N \Delta}$, of 700, 800, 900
and 1000 MeV, and employing the NA24 pion structure function
\cite{glue92}. All curves shown correspond to a four-momentum
transfer of $Q^2=1$ GeV$^2$, and, where not directly available, the
various PDFs were evolved by numerically \cite{koba94} solving the
non-singlet QCD evolution equations \cite{alta77}.

\figs{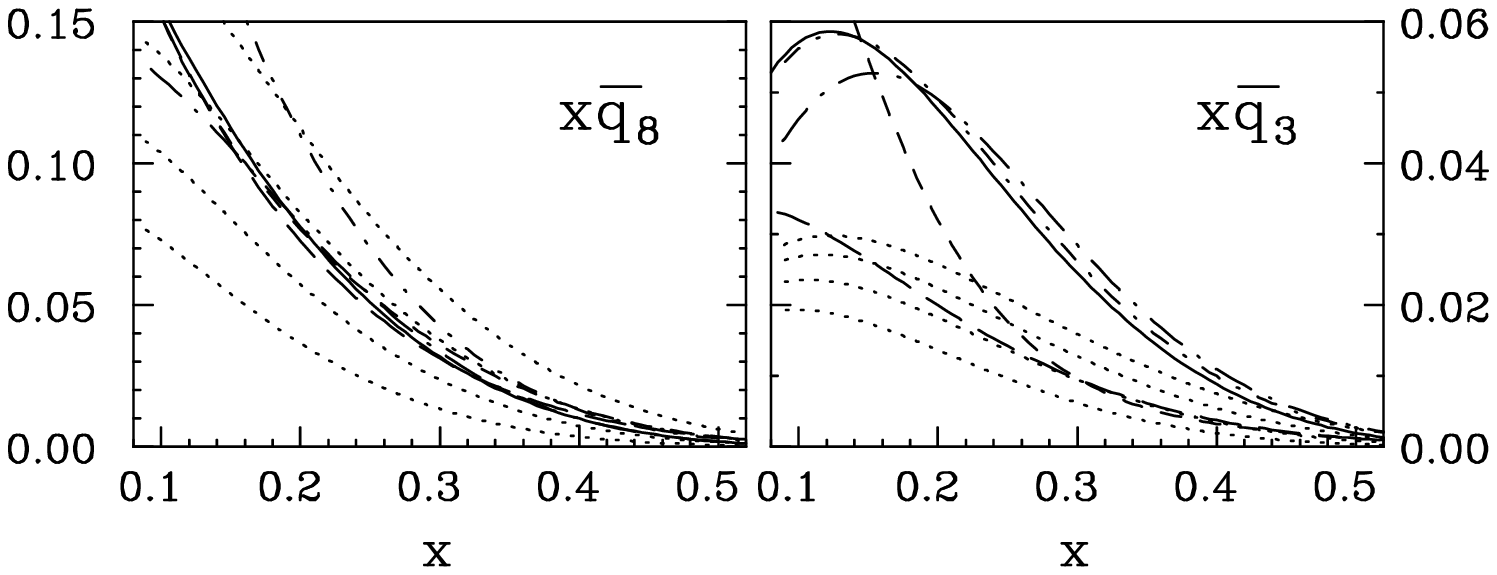}{ht}{7.5}{0.15}{2.3}{0.5}{The flavor
breaking components of the nucleon's antiquark sea at $Q^2=1$
GeV$^2$. The solid [dot-long-dashed, dot-short-dashed, short-dashed,
long-dashed] curves correspond to the MRS(A') [CTEQ3M, GRV94(HO),
BM(A), MRS(D'$_-$)] PDFs, and the dotted lines refer to our meson
cloud calculations using an exponential form factor and varying
$\Lambda_e$ between 700 and 1000 MeV.}

As can be seen from Fig.~2, the various parton distribution functions
(PDFs) agree well with each other in the $SU(3)$ channel, while there
are significant discrepancies for the $SU(2)$ breaking component.
Note that in a recent analysis of neutrino charm production it was
observed that the nucleon's strange quark content is suppressed
with respect to its non-strange sea by a factor of $\kappa=0.48\pm
0.06$ \cite{ccfr95}. This is consistent with the two most recent PDFs
of Refs.~\cite{mart95} and \cite{cteq94}, where this reduction factor
is $\kappa = 0.5$. Also, these two PDFs agree well with the
$x$-dependence of the $(\bar d - \bar u)$ asymmetry, as measured by
the NA51 collaboration at CERN \cite{bald94} and the CDF
collaboration at the FNAL $p\bar p$ collider \cite{fnal92}, while, for
instance, in the BM(A) and MRS(D'$_-$) distributions this asymmetry is
concentrated at smaller $x$-values.

It is obvious from Fig.~2, that we can obtain a satisfactory fit to
the $SU(3)$ breaking share of the nucleon's antiquark distribution,
$x \bar q_8(x,Q^2)$, by varying the cut-off parameter in the
aforementioned range. However, the agreement with the $SU(2)$ breaking
component, $x \bar q_3(x,Q^2)$, this fit will yield, will be rather
poor. In Table I, we nevertheless list the results of such a fit to
the already shown and some additional parton distribution functions
used previously in this context. We employ two different pion
structure functions, the NA10 parametrization from Ref.~\cite{sutt92}
and the NA24 parametrization from Ref.~\cite{glue92}, fit to
Drell-Yan data, and compare with the PDFs in the range of $x$-values
of $0.2 \le x \le 0.5$. Again, where not directly
available, the various PDFs were evolved down to the desired $Q^2$
by employing the LAG2NS code from Ref.~\cite{koba94}.
The corresponding results are displayed in Fig.~3 for the MRS(A') and
CTEQ3M parametrizations for values of the four-momentum transfer
of 1 and 4 GeV$^2$.

\begin{table}[hbt]
\begin{tabular}{lcc@{~~~~}|dd@{~~~~}|dd@{~~~~}|dd@{~~~~}}
PDF & Ref. & Year & \multicolumn{6}{c}{$\Lambda^e_{\pi NN}=
\Lambda^e_{KNY}=\Lambda^e_{\pi N\Delta}$ \ [MeV]}             \\
\hline
    &   &  & \multicolumn{2}{c@{~~~~}|}{$Q^2=1$ GeV$^2$}
           & \multicolumn{2}{c@{~~~~}|}{$Q^2=2$ GeV$^2$}
           & \multicolumn{2}{c}{$Q^2=4$ GeV$^2$}                    \\
    &   &  & NA10 & NA24 & NA10 & NA24 & NA10 & NA24                \\
\hline
MRS(A')    &\protect\cite{mart95}&(1995)&820&870&810&870&810&860    \\
CTEQ3M     &\protect\cite{cteq94}&(1994)&830&880&810&870&810&860    \\
GRV94(HO)  &\protect\cite{glrv94}&(1994)&900&940&890&950&890&940    \\
BM(A)      &\protect\cite{berg93}&(1993)&830&880&830&880&820&880    \\
MRS(D'$_-$)&\protect\cite{mart93}&(1993)&810&860&810&860&810&860    \\
GRV(HO)    &\protect\cite{glu92a}&(1992)&800&850&790&840&790&840    \\
MT(NS)     &\protect\cite{morf91}&(1991)&700&730&690&730&690&730    \\
HMRS(E)    &\protect\cite{harr90}&(1990)&730&770&740&790&750&790    \\
DFLM       &\protect\cite{diem88}&(1988)&770&810&780&830&790&850    \\
EHLQ(I)    &\protect\cite{eich84}&(1984)&680&710&680&720&680&720    \\
\end{tabular}
\vskip 0.15in
\caption{Results of a fit to the flavor $SU(3)$ breaking component of
the nucleon's antiquark distribution at various $Q^2$ and for
different PDFs. The range of $x$-values considered is $0.2 \le x \le
0.5$.}
\end{table}

\figs{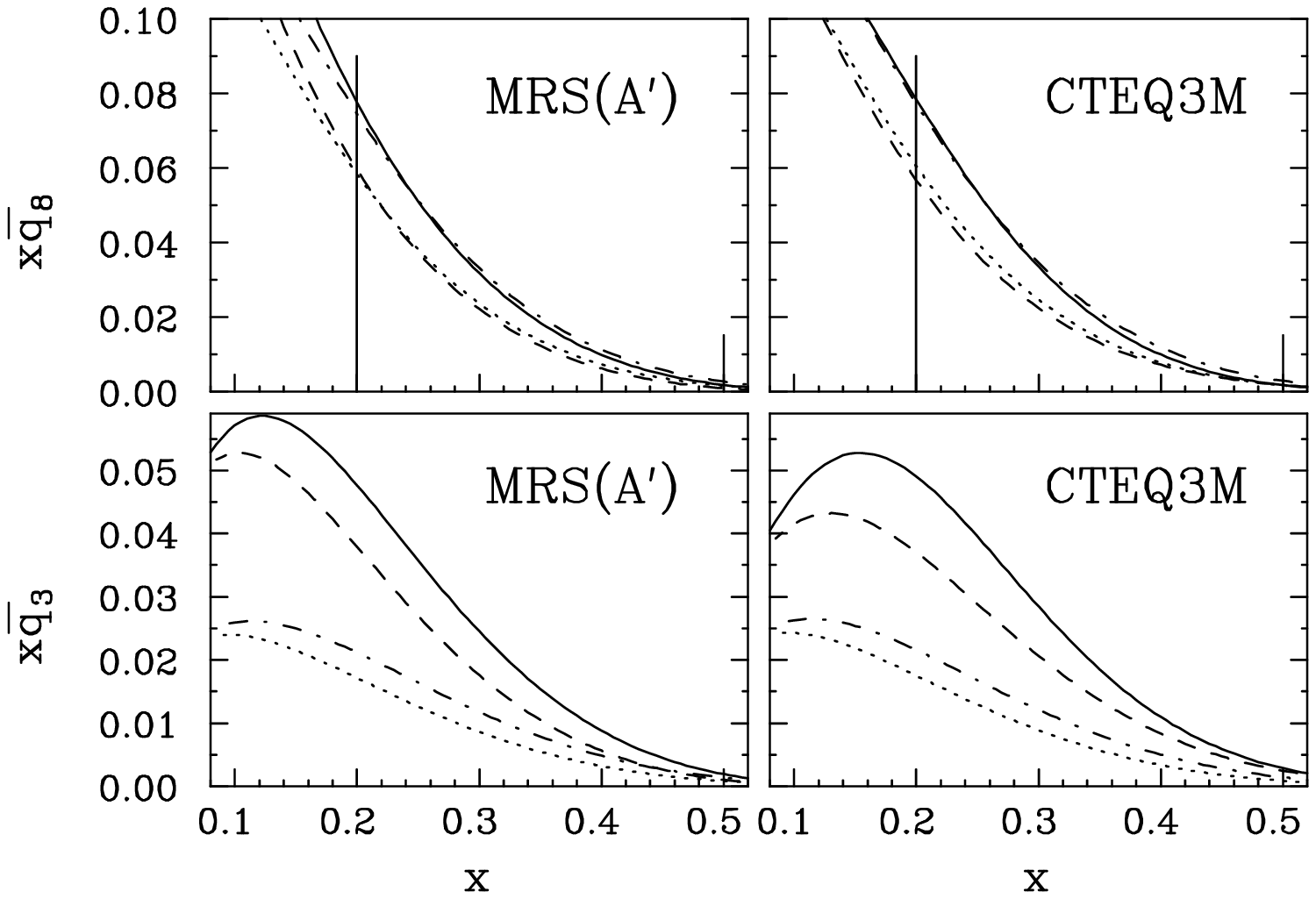}{htb}{7.5}{-0.55}{4.1}{0.75}{The flavor
breaking components of the nucleon's antiquark sea. The solid
and dashed curves show the PDFs at $Q^2=1$ and 4 GeV$^2$, and the
dot-dashed and dotted lines refer to the corresponding meson cloud
calculations using the NA24 pion structure function and employing
Gaussian cut-offs of $\Lambda_e=870$ and 880 MeV for the MRS(A') and
CTEQ3M PDFs, respectively. The range of $x$-values considered in our
fits is indicated through the vertical lines.}

The MT \cite{morf91}, HMRS \cite{harr90} and EHLQ \cite{eich84}
parametrizations, which would suggest even softer cut-offs,
are today assumed to underestimate the nucleon's sea quark content.
In the very recent GRV94 parametrization \cite{glrv94}, on the other
hand, it is assumed that the nucleon's strange quark sea vanishes at
a very low renormalization scale, whereas $x \bar s$ is only
suppressed by a factor of $\kappa=0.5$ in the PDFs of
Refs.~\cite{mart95} and \cite{cteq94}. This manifests itself in
a smaller strange quark sea, and hence a larger $x \bar q_8$,
for the GRV94 parametrization, as can be seen in Fig.~2, and it
leads to larger values for the cut-off parameter $\Lambda^e_{\pi NN}$
-- but to softer form factors at the $KNY$ vertex -- that
are necessary to describe this PDF. Furthermore, the $SU(3)$ breaking
components of the MRS(A'), CTEQ3M, BM(A) and MRS(D'$_-$)
distributions agree quite well with each other, and the major
uncertainties in our fit arise through the differences between the
NA10 and NA24 pion structure functions.

Averaging over the MRS(A') and CTEQ3M parton distribution functions
and the NA10 and NA24 pion structure functions at a scale of $Q^2=1$
GeV$^2$ we find an exponential cut-off mass of
\begin{equation}
\Lambda_e = \Lambda^e_{\pi NN} = \Lambda^e_{KNY} =
\Lambda^e_{\pi N \Delta} \approx 850~\mbox{MeV}
\ .
\label{eq:r1d}
\end{equation}

This is significantly larger than the values\footnote{The results of
the various calculations were translated into exponential form
employing Eq.~(\ref{eq:kum}).} given by Melnitchouk and Thomas
\cite{meln93} (650 MeV), by Frankfurt et al.~\cite{fran89} (700 MeV),
by Thomas \cite{thom83} (700 MeV), and by Kumano \cite{kuma91}
(750 MeV), where the now outdated MT \cite{morf91}, HMRS
\cite{harr90}, DFLM \cite{diem88}, EHLQ \cite{eich84}, DO
\cite{owen91} or FF \cite{fiel77} parton distribution functions were
used. It approximately agrees with the result of Melnitchouk et
al.~\cite{meln91} (800 MeV), yet it is smaller than the
values obtained by the J\"ulich group \cite{hwan91,szcz93} (950 MeV)
and by Signal et al.~\cite{sign91} (1000 MeV). The reason for the
latter discrepancies is, firstly, that in Refs.~\cite{hwan91,szcz93}
an additional renormalization of the Sullivan process was used
which diminishes the mesonic contribution and hence leads to larger
cut-offs and, on the other hand, that in Ref.~\cite{sign91} the
$N\to\Delta\pi$ process was treated non-relativistically.

As can be seen from Fig.~3, our meson cloud calculations reasonably
describe the $SU(3)$ breaking share of the nucleon's
antiquark sea at the scale where the vertices were
originally fit, i.e., at $Q^2=1$ GeV$^2$. However, also at a
scale of $Q^2=4$ GeV$^2$, our calculations match the PDFs in
that channel quite well. This suggests that for
this component the perturbative evolution of the parton
distributions in the nucleon as well as the pion by means of the
non-singlet QCD evolution equations is compatible with
the convolution of the latter distributions.

\subsection{The nucleon's full antiquark sea:}

It is obvious from Fig.~3 that our description of the flavor
$SU(2)$ breaking component, $x \bar q_3(x,Q^2)$, is very poor, with
the meson cloud calculations underestimating the data by about a
factor of 2. We could cure this discrepancy, while preserving
the good agreement in the $SU(3)$ channel, by allowing the
$\pi NN$, $KNY$ and $\pi N\Delta$ vertices to be
different, as is evident from Eqs.~(\ref{eq:su3a}) and
(\ref{eq:su2a}). In the following we thus vary
the cut-offs in the octet and decuplet channel and in the strange
and non-strange sectors separately. In essence, we fit
$\Lambda^e_{KNY}$ to the nucleon's strange quark content,
$x \bar s(x,Q^2)$, and we then adjust
$\Lambda^e_{\pi NN}$ and $\Lambda^e_{\pi N \Delta}$ to get an optimal
description of the $SU(3)$ and $SU(2)$ flavor breaking components of
the nucleon's antiquark sea, $x \bar q_8(x,Q^2)$ and $x \bar
q_3(x,Q^2)$.

As the strange sea, $x \bar s(x,Q^2)$, contains a significant flavor
singlet component, we are now {\em forced}
to evaluate the PDFs at a small value of the four-momentum
transfer where the contamination from gluon splitting into $q\bar q$
pairs is not yet dominant and where a comparison of hadronic and
quark-gluon degrees of freedom might still be possible. In
particular, we choose to work at a scale of $Q^2=1$ GeV$^2$ -- but
we will also show results for larger $Q^2$ -- and we consider
$x$-values in the range of $0.3 \le x \le 0.5$.
The results of such fits to the two most recent PDFs of
Refs.~\cite{mart95} and \cite{cteq94} are listed in Table II.
As the MRS(A') parametrization of Ref.~\cite{mart95} is not yet
available at such low $Q^2$, we restrict ourselves from now on to
the very similar MRS(A) \cite{mart94}
parton distribution function. Furthermore, in
the singlet channel also the meson's {\em sea quark} distributions
contribute, however, at quite small $x$-values only.

\begin{table}[hbt]
\begin{tabular}{lc@{~~~~}|dd@{~~~~}|dd@{~~~~}|dd@{~~~~}}
PDF & $Q^2$ [GeV$^2$]
    & \multicolumn{2}{c@{~~~~}|}{$\Lambda^e_{\pi NN}$\ [MeV]}
    & \multicolumn{2}{c@{~~~~}|}{$\Lambda^e_{\pi N\Delta}$\ [MeV]}
    & \multicolumn{2}{c}{$\Lambda^e_{KNY}$\ [MeV]}            \\
    &      & NA10 & NA24 & NA10 & NA24 & NA10 & NA24                \\
\hline
MRS(A)      \protect\cite{mart94}& 1 & 960&1010&780&790&1150&1180   \\
                                 & 2 & 970&1030&800&820&1180&1220   \\
                                 & 4 & 970&1030&840&860&1210&1250   \\
\hline
CTEQ3M      \protect\cite{cteq94}& 1 & 990&1050&810&810&1180&1210   \\
                                 & 2 &1010&1070&820&840&1220&1260   \\
                                 & 4 &1010&1080&860&870&1250&1290   \\
\end{tabular}
\vskip 0.15in
\caption{Results of a fit to $x \bar s(x,Q^2)$, $x \bar q_8(x,Q^2)$
and $x \bar q_3(x,Q^2)$ of the nucleon at various scales, $Q^2$, of 1,
2 and 4 GeV$^2$. The range of $x$-values under consideration is $0.3
\le x \le 0.5$.}
\end{table}

Again averaging over the MRS(A) and CTEQ3M parton distributions
and the NA10 and NA24 pion structure functions we find exponential
cut-offs of
\begin{mathletters}
\begin{eqnarray}
\Lambda^e_{\pi NN}      & \approx & ~1000~\mbox{MeV} \ ,
\label{eq:r31}
\\
\Lambda^e_{\pi N\Delta} & \approx & ~~800~\mbox{MeV} \ ,
\label{eq:r32}
\\
\Lambda^e_{KNY}   & \approx & ~1180~\mbox{MeV} \ ,
\label{eq:r33}
\end{eqnarray}
\end{mathletters}
in the various relevant meson-baryon sectors at a scale of $Q^2=1$
GeV$^2$. The corresponding antiquark
distributions are depicted in Fig.~4 for values of the four-momentum
transfer of 1 and 4 GeV$^2$. Also shown are a few data points for
$x \bar s(x,Q^2)$ as given by the CCFR collaboration \cite{ccfr95}.

\figs{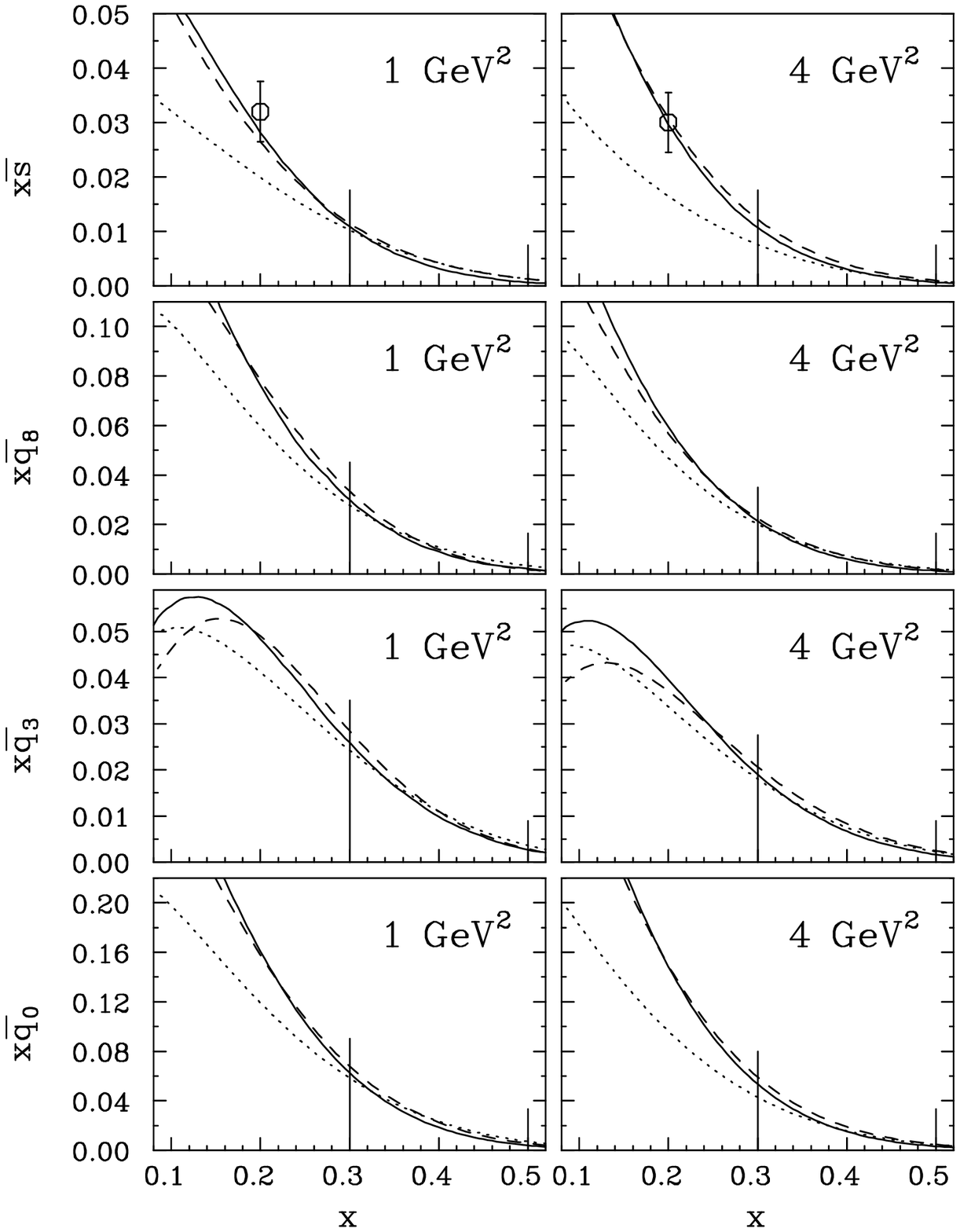}{p}{7.5}{-0.55}{7.7}{1.30}{The various components of
the nucleon's antiquark sea. The solid and dashed curves show the
MRS(A) and CTEQ3M PDFs, and the dotted lines refer to our meson cloud
calculations using the NA24 pion structure function and the cut-offs
$\Lambda^e_{\pi NN}=1030$ MeV, $\Lambda^e_{\pi N\Delta}=800$ MeV and
$\Lambda^e_{KNY}=1200$ MeV that were adjusted at a scale of
$Q^2=1$ GeV$^2$.}

As can be seen from Fig.~4, allowing the $\pi NN$ and
$\pi N\Delta$ vertices to be different significantly improves the
quality of our fits to the $SU(2)$ flavor breaking component, $x \bar
q_3(x,Q^2)$. Also, the agreement in both flavor breaking channels,
$x \bar q_8(x,Q^2)$ and $x \bar q_3(x,Q^2)$, extends to smaller
$x$-values than those actually considered in our fits, and it remains
satisfactory when the four-momentum transfer is increased. This shows
that the meson cloud picture put forward in this work may yield a
satisfactory description of the nucleon's {\em flavor
non-singlet} anitquark distributions down to $x$-values of
$x \gtrsim 0.2$. Also, the perturbative evolution of the partonic
distributions in the nucleon as well as the pion seems to be
compatible with the convolution of the latter quantities in
the non-singlet channels.

The situation is very different for the {\em flavor singlet}
component. Due to the mixing with gluonic degrees of freedom, most
notably through gluon splitting into $q\bar q$ pairs, we only find
reasonable agreement with the tails of the $x \bar s$ and $x \bar q_0
\equiv x[\bar u + \bar d + \bar s]$ PDFs in the region of $x \gtrsim
0.3$. In addition, neglected shadowing effects tend to
further decrease the cross section of lepton scattering from the $MB$
component of the nucleon's wave function at small $x$.
Furthermore, for $x \bar s$ and $x \bar q_0$ the deviations
between our meson cloud calculations and the data-based
parametrizations grow rapidly with increasing $Q^2$, as can also be
seen from Fig.~4. This indicates
that, in the singlet channel, the evolution of the parton distribution
functions is incompatible with the meson cloud convolution picture
investigated here. This stresses the importance of gluonic degrees of
freedom in the flavor singlet channel, and it shows that the virtual
meson cloud alone is not able to yield the nucleon's entire antiquark
sea, even at a low scale of $Q^2=1$ GeV$^2$.

\section{DISCUSSION}

\subsection{The meson-nucleon vertices:}

The vertices that were determined in this work are still significantly
softer than what is used in most meson-exchange models of the $NN$
interaction or in evaluations of corresponding meson-exchange
currents.\footnote{In most MEC calculations the one-pion exchange is
treated with point-like vertices combined with a hard-core
short-distance cut-off in coordinate space, and, at this point, it is
unclear whether this is consonant with our findings.}
If we translate the $\Lambda^e_{\pi NN}$ of Eq.~(\ref{eq:r31}) into a
monopole cut-off using the approximate relation of
Eq.~(\ref{eq:kum})\footnote{We emphasize again that the
contribution from decuplet baryons diverges when a monopole form
factor is used, and that, hence, Eq.~(\ref{eq:kum}) is only applied in
a qualitative manner in order to be able to compare with other works.}
we find a value of $\Lambda^m_{\pi NN} \approx 800$ MeV. This is much
smaller than the respective quantity, $\Lambda_{\pi NN}=1.3$ GeV,
employed in the framework of the Bonn potential \cite{mach87}, or the
$\Lambda_{\pi NN}=1.2$ GeV used in the evaluation of respective
meson-exchange currents \cite{stru87}.  On the other hand,
the $\Lambda^e_{\pi NN}$ of Eq.~(\ref{eq:r31}) is not too far from
the exponential cut-off used in the corresponding channel in the
Nijmegen potential \cite{nijm91}, $\Lambda_{PV} = 1195$ MeV, or the
Gaussian cut-off parameter employed by van Kolck et al.
\cite{kolc94}, $\Lambda \approx 1100$ MeV, in their evaluation of
nuclear forces from a Chiral Lagrangian.

In Ref.~\cite{muso94} kaon loop contributions to low-energy strange
quark matrix elements, which will eventually be measured at CEBAF
\cite{cebafp}, were modelled employing the vertices from the Bonn YN
potential \cite{holz89}. Note that the corresponding $KNY$
cut-offs are significantly harder than that of Eq.~(\ref{eq:r33}),
which, in turn, has been determined directly from the
strange quark distribution in the nucleon. Thus, if alternatively the
quantity of Eq.~(\ref{eq:r33}) would be used in their analysis
the respective strangeness matrix elements would be strongly reduced,
most likely down to the point where they could no longer be observed
at CEBAF.

The cut-offs in Eqs.~(\ref{eq:r31}) and (\ref{eq:r32}) indicate that
the $\pi N\Delta$ vertex is considerably softer than the
$\pi NN$ vertex.\footnote{It was also mentioned in Ref.~\cite{hwan91}
that better agreement with the NMC data could be achieved in their
meson cloud calculation if the $\pi N \Delta$ vertex would be
softer than the $\pi NN$ vertex.} This is in qualitative
agreement with the faster fall-off of the electromagnetic $p
\to \Delta$ transition form factor relative to the nucleon's elastic
e.m.~form factor as observed in exclusive resonance electron
scattering \cite{stol93}. Only in a pure $SU(6)$ model,
where the quarks' spatial wave functions in the nucleon as well as
the $\Delta$ are all identical $s_{1/2}$ states, are the form factors
governing the processes $N \to \pi N$ and $N \to \pi\Delta$
necessarily the same. Already the one-gluon-exchange color-magnetic
hyperfine interaction breaks that symmetry, and introduces a small
$L=2$ contribution into the $\Delta$ ground state \cite{isgu78}.
There is experimental evidence for that admixture -- and hence the
breaking of the na\"\i ve $SU(6)$ symmetry -- through the
non-vanishing E2/M1 electromagnetic transition ratio observed in
photoproduction data \cite{davi90}.

\subsection{Structure of the nucleon:}

Using a relation given by Thomas \cite{thom83}, the $\Lambda^e_{\pi
NN}$ of Eq.~(\ref{eq:r31}) can be converted into a MIT bag radius of
$R \approx 0.60$ fm. This is somewhat smaller than what is usually
cited in the literature \cite{MITmod} ($R \approx 1$ fm) when the
nucleon's standard low-energy observables, as, for instance,
its charge radius and magnetic moment, are described in the framework
of that model. However, it agrees with the value favored by the
Adelaide group \cite{sign88} for the evaluation of the nucleon's
deep inelastic structure functions from the MIT bag model. Our
results, therefore, indicate a characteristic confinement radius of
the quarks in a nucleon of about 0.6 fm at a scale of $Q^2
\approx 1$ GeV$^2$. Note, however, that the pion cloud yields a
significant contribution to the nucleon's charge radius, i.e.,
$\langle r_c^2 \rangle = \langle r_q^2 \rangle + \langle r_\pi^2
\rangle$, and there is thus no contradiction between the result given
here and the proton's experimental charge radius of 0.86 fm
\cite{simo80}.

Furthermore, the perturbative two-phase picture of the nucleon, that
was applied in this investigation, only makes sense if the mesonic
component is just a perturbation, i.e., if the vertices are not too
hard. Otherwise, higher-order diagrams with more than one pion present
grow important, and the convergency of the whole approach becomes
questionable. A quantitative measure of that is, for instance, the
"number of mesons in the air",
\begin{equation}
n_M = \int_0^1 dy f_M(y)
\ ,
\label{eq:npi}
\end{equation}
which is $n_\pi = 0.66$ and $n_K = 0.10$ for the cut-offs of
Eqs.~(\ref{eq:r31}) to (\ref{eq:r33}).
Those values are already quite close to this limit of
applicability of models of that type.

In Fig.~5, we visualize the dominant contributions to the convolution
integrals of Eqs.~(\ref{eq:su3a}) and (\ref{eq:su2a}). In particular,
it is analyzed which region in the space spanned by the mesons'
light-cone momentum fraction $y$, its virtuality $t$ as well as
the loop momentum $|{\bf k}|$ yields the largest share to $x \bar
q_8(x,Q^2)$ for a typical $x$-value of $x=0.3$ and for the case of
the $N \to N\pi$ sub-process. We used a Gaussian cut-off of
$\Lambda^e_{\pi NN}=1030$ MeV and employed the NA24 pion structure
function at a scale of $Q^2=1$ GeV$^2$.

\figs{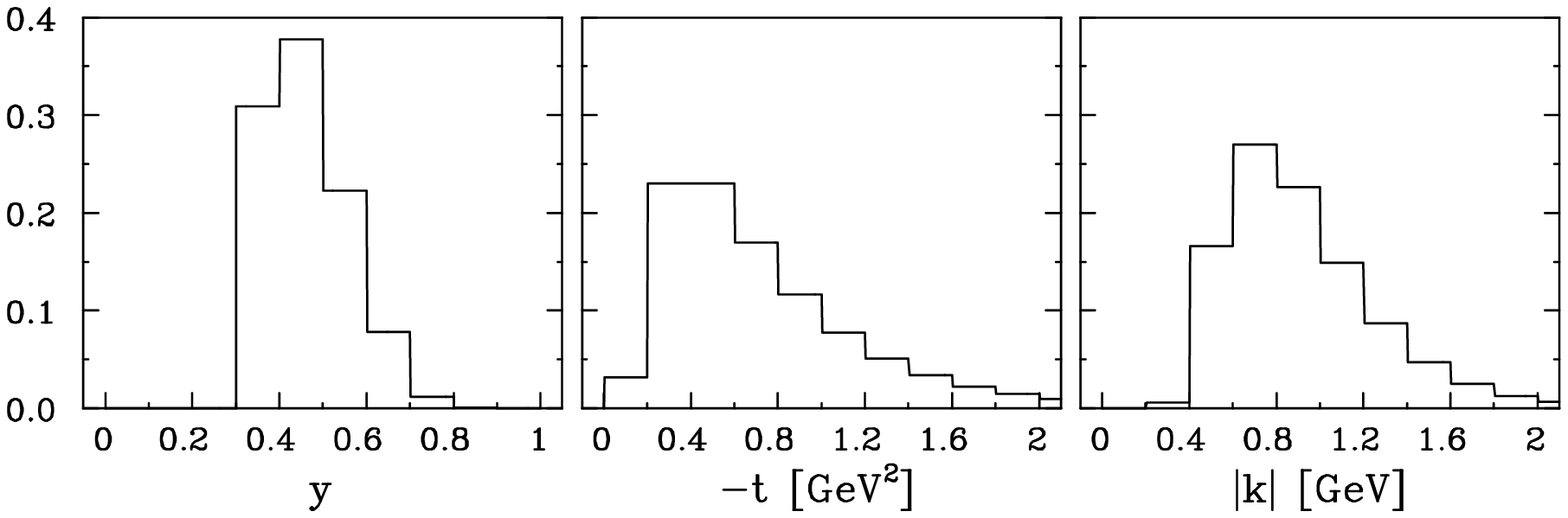}{htbp}{6.75}{-0.22}{2.15}{0.05}{The different relative
contributions to the convolution integral of
Eq.~(\protect\ref{eq:su3a}) for the $N \to N\pi$ process
from the various regions in the space spanned by the quantities
$y$, $t$ and $|{\bf k}|$, and for a typical $x$-value of $x=0.3$.}

Fig.~5 shows that, for $x=0.3$, the most relevant light-cone momentum
fraction is the region $0.3 \leq y \leq 0.5$, which contributes 70\%
to the integral. Typically, the mesons in the loop are also highly
virtual, with the areas $0.2~\mbox{GeV}^2 \leq -t \leq
0.6~\mbox{GeV}^2$, $0.6~\mbox{GeV}^2 \leq -t \leq 1.0~\mbox{GeV}^2$
and $-t \geq 1.0~\mbox{GeV}^2$ yielding 45\%, 30\% and 20\%,
respectively. And, in contrast to low-energy nuclear physics
phenomena where meson momenta around the pion mass are probed, the
relevant loop momenta are here of the order of $|{\bf k}| \approx
0.8$ GeV. In detail, the region of $|{\bf k}| \leq 0.5$ GeV, which
is most important in $NN$ meson-exchange potentials, contributes only
15\% to the deep inelastic process considered here. The dominant
contributions, on the other hand, stem from the areas $0.5~\mbox{GeV}
\leq |{\bf k}| \leq 1.0~\mbox{GeV}$ and $|{\bf k}| \geq
1.0~\mbox{GeV}$, which yield 50\% and 30\%, respectively. This
indicates i) that a comparison of the cut-offs that we
determine here with the corresponding quantities used in low-energy
nuclear physics has to be looked upon with some caution, and ii) that
the very idea of using meson clouds at such virtualities seems quite
doubtful.

\subsection{Gottfried sum rule:}

If we evaluate the meson cloud contribution to the Gottfried sum rule,
\begin{mathletters}
\begin{eqnarray}
S_G & \equiv & {1 \over 3} - {2 \over 3} \Delta_{G} \equiv
\int^1_0 dx \, {F_2^{\mu p}(x,Q^2) - F_2^{\mu n}(x,Q^2) \over x} \ ,
\label{eq:go1}
\\
\noalign{\medskip}
& \equiv & {1 \over 3} \int^1_0 dx \,
           [u(x,Q^2) + \bar u(x,Q^2) - d(x,Q^2) - \bar d(x,Q^2)] \ ,
\label{eq:go2}
\\
\noalign{\medskip}
& = & {1 \over 3}~+~{2 \over 3} \int^1_0 dx \,
                     [\bar u(x,Q^2) - \bar d(x,Q^2)] \ ,
\label{eq:go3}
\end{eqnarray}
\end{mathletters}
with the vertices of Eqs.~(\ref{eq:r31}) to (\ref{eq:r33}), we find
a value of $\Delta_G = 0.17$ at $Q^2=4$ GeV$^2$. This agrees, within
the error-bars, with the quantity given by the NMC collaboration of
$\Delta_G^{exp} = 0.148 \pm 0.039$ \cite{amau91}. Note that the
direct meson cloud diagram, depicted in Fig.~1a, yields no
contribution to Eq.~(\ref{eq:go2}) due to $G$ parity. It has been
argued, for instance in Ref.~\cite{sign91}, that the entire effect
is thus due to the recoil baryon being struck by the incoming meson,
i.e., the diagram shown in Fig.~1b. However, conversely, only the
direct meson cloud diagram of Fig.~1a contributes to
Eq.~(\ref{eq:go3}). The obvious solution to this paradox \cite{szcz93}
is that Eqs.~(\ref{eq:go2}) and (\ref{eq:go3}) are related through the
normalization requirements for the valence quark distributions,
\begin{mathletters}
\begin{eqnarray}
\int^1_0 dx \, [u(x,Q^2) - \bar u(x,Q^2)] & = & 2 \ ,
\label{eq:go5}
\\
\int^1_0 dx \, [d(x,Q^2) - \bar d(x,Q^2)] & = & 1 \ ,
\label{eq:go6}
\end{eqnarray}
\end{mathletters}
and the distinction between the direct mesonic and the recoil
diagram     is thus artificial, at least for the meson cloud
contribution to the violation of the Gottfried sum rule.

It is common wisdom that, at small $x$ and high energies,
the dependence of the parton distributions on the invariant mass
squared, $W^2$, is governed by the exchange of the leading
Regge pole in the $t$-channel of the elastic amplitude, which for the
$SU(2)$ flavor breaking share of the nucleon's antiquark distribution
leads to
\begin{equation}
x[\bar d(x,Q^2)-\bar u(x,Q^2)]\Big|_{x<0.05} = c_3~x^{1-\alpha(A_2)}
\ ,
\label{eq:reg}
\end{equation}
where $\alpha(A_2) \approx 0.4$ for the $A_2$-meson's Reggeon that is
relevant in this channel. In Ref.~\cite{strikm} it was argued that
due to nuclear shadowing in the deuteron the cross section ratio
$F_{2n}/F_{2p}$ which was measured by the NMC collaboration
\cite{nmc291} has to be modified at small $x$, and that hence
$\bar d(x) / \bar u(x) > 1$ at $x=0.007$. This indicates that the
constant $c_3$ in Eq.~(\ref{eq:reg}) is larger than that assumed in
the current parametrizations,
and it was shown in Ref.~\cite{strikm} that the deviation of the
Gottfried integral from $1/3$ observed by the NMC collaboration --
which, in addition, should also be somewhat bigger than the value
$\Delta_G^{exp}$ given in Ref.~\cite{amau91} again due to nuclear
shadowing effects in the deuteron -- can already be saturated by the
region $0 < x < 0.02$ in the integral in Eq.~(\ref{eq:go3}).

This suggests a reduction of the quantity $x(\bar d - \bar u)$ at
moderate $x$, as was originally assumed, for
instance, in the MRS(D'$_-$) PDF, and it would
lead to a decrease of $\Lambda^e_{\pi NN}$ of
Eq.~(\ref{eq:r31}) and to an increase in $\Lambda^e_{\pi N\Delta}$
of Eq.~(\ref{eq:r32}) in the realm of the meson cloud picture.
Note, also, that only about 30\% of the integral in
Eq.~(\ref{eq:go3}) stems from the area of $x > 0.1$, where the meson
cloud picture is at least somehow applicable, and the dominant
contribution to $\Delta_G$ comes from the small-$x$ region. This
underlines our claim that, in this context, it is more appropriate to
study the tails of the antiquark distributions than to solely
concentrate on sum rules. There is an approved
experiment at Fermilab, E866, which will measure the quantity
$x(\bar d - \bar u)$ over the whole relevant kinematic region
and which will settle this issue \cite{e86692}.

\subsection{Scale dependence and small x physics:}

As already mentioned in the last section, the description of the {\em
flavor non-singlet} share of the nucleon's antiquark distributions is
quite satisfactory for $x$-values larger than about 0.2, and the
quality of the agreement of our meson cloud calculations with the
data-based PDFs is independent of the scale $Q^2$. This indicates
that the evolution of the partonic distributions in the framework of
perturbative QCD and the convolution picture are compatible with
each other, at least in the non-singlet channels.

The {\em flavor singlet} component, on the other hand, is quite
poorly approximated through the mesonic cloud, even at a small
scale of $Q^2=1$ GeV$^2$.  We are, in fact, only able to attribute
the tails of the $x \bar q_0$ and $x \bar s$
distributions -- the area of $x \gtrsim 0.3$ -- to the nucleon's
virtual meson cloud, and the deviations from the data-based
parametrizations grow rapidly with increasing $Q^2$. This indicates
that, in the singlet channel, other degrees of freedom, most notably
gluon splitting into $q\bar q$ pairs, are relevant, even at moderate
$x$ and small values of the four-momentum transfer.

The scale dependence of our results is analyzed in greater detail
in Fig.~6, which depicts the $\pi NN$, $\pi N\Delta$ and $KNY$
vertices that were determined in Sect.~III.B (see Table II) from fits
to the tails of the nucleon's entire antiquark sea at different
values of the four-momentum transfer.

\figs{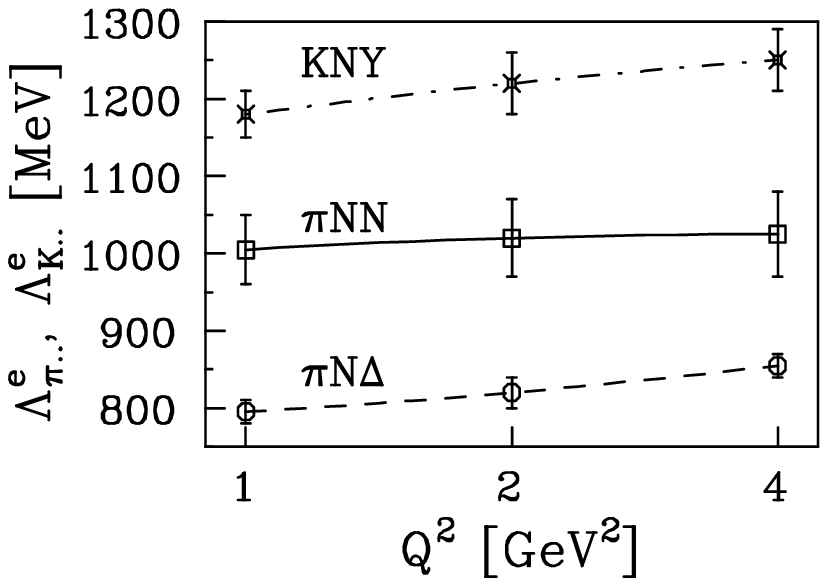}{htb}{7.5}{-0.6}{2.35}{2.25}{The cut-off parameters
$\Lambda^e_{\pi NN}$, $\Lambda^e_{\pi N\Delta}$ and
$\Lambda^e_{KNY}$ adjusted to fit the nucleon's antiquark
distributions as functions of the scale, $Q^2$, at which this fit is
performed.}

As can be seen from Fig.~6, the parameter $\Lambda^e_{KNY}$
that is determined from $x \bar s(x,Q^2)$ which, in turn, is mostly
flavor singlet shows a very strong scale dependence, and
$\Lambda^e_{\pi NN}$ which is fixed through the flavor non-singlet
distributions $x \bar q_8(x,Q^2)$ and $x \bar q_3(x,Q^2)$ hardly
changes with $Q^2$. This substantiates what has been discussed in the
above about the qualitatively different behavior in the flavor
singlet and non-singlet channels. The variations of $\Lambda^e_{\pi
N\Delta}$ with $Q^2$, on the other hand, are induced via the
interplay of Eqs.~(\ref{eq:su3a}) and (\ref{eq:su2a}) through the
scale dependence of the $KNY$ vertex which contributes not
only to $x \bar s(x,Q^2)$ but also to the flavor non-singlet
distribution $x \bar q_8(x,Q^2)$.

The strong $Q^2$ dependence of the $KNY$ vertex signals
a correlation between the presence of gluonic degrees of freedom
at low $Q^2$ and contributions from the nucleon's virtual kaon cloud.
At this point, both processes are indistinguishable, and the
$\Lambda^e_{KNY}$ given here has to be understood as an {\em
upper bound}. As, in turn, a softer $KNY$ vertex would lead
also to a softer $\pi NN$ vertex, as can be inferred from
Eq.~(\ref{eq:su3a}), the values for $\Lambda^e_{\pi NN}$ listed in
this work are actually {\em upper bounds} on that quantity.
Thus, the scenario presented here has to be
viewed as an extreme case. This conjecture is
further supported by the fact that the meson cloud can only
describe the tail of the flavor singlet antiquark distribution,
while the mismatch at $x \lesssim 0.3$ is probably due to gluon
splitting into $q\bar q$ pairs, and there are no stringent reasons
why the influence of gluonic degrees of freedom should be
limited to the aforementioned small-$x$ range only.

This suggests that a smooth transition between the low-energy
nuclear physics degrees of freedom, i.e., baryons and mesons, and
a description in terms of quarks and gluons which is adequate for
hard processes does not seem to exist. It has been clearly shown in
this work that, even at a small $Q^2$, it is {\em not} possible to
attribute the entire   sea quark distributions in the nucleon to its
mesonic cloud.\footnote{In this conjecture we manifestly differ with
Refs.~\cite{hwan91,szcz93}.} Firstly, the meson-nucleon vertices
which are necessary to describe the {\em flavor breaking} share of
the nucleon's antiquark distributions in the meson cloud picture
outlined here are significantly different from the respective
quantities used in low-energy nuclear physics, as, for instance, the
cut-offs employed in the Bonn potential. And, furthermore, even when
freely adjusting the respective form factors, a satisfactory
description of the {\em flavor singlet} component of the nucleon's
antiquark sea is not possible. Not only are there large discrepancies
already for $x$-values smaller than about 0.3, but also the
perturbative evolution of the singlet component to higher values of
the four-momentum transfer by means of the QCD evolution equations
is incompatible with the meson cloud convolution picture.
This shows that gluon degrees of freedom play an important role
in the structure of the nucleon, even at a scale of $Q^2=1$ GeV$^2$.

At small $x$ and high energies the virtual photon converts
into a hadronic $q \bar q$ state at a distance
\begin{equation}
l \approx {1 \over 2 m_N x}
\label{eq:sh1}
\end{equation}
in the target rest frame.
If this coherence length is larger than the dimension of the target,
\begin{equation}
l \gtrsim 2 \, \langle r_T \rangle
\ ,
\label{eq:sh2}
\end{equation}
it is not the virtual photon probing the target but this
$q \bar q$ state, and this is the basis of the shadowing phenomenon.
Hence, the na\"\i ve impulse approximation picture that was applied
in this work to the nucleon's virtual meson cloud is only applicable
if the relevant distances are larger than the coherence length $l$.
As the important meson loop momenta are of the order of $|{\bf k}|
\approx 0.8$ GeV, the significant distances between the nucleon and
the mesons driving the convolution integrals are $\langle r_{MN}
\rangle \approx 0.25$ fm, and the shadowing condition in
Eq.~(\ref{eq:sh2}) is already satisfied at $x \lesssim 0.2$. This
underlines our conclusion that at small $x$, where absorptive effects
are important, the meson cloud picture is not adequate and different
degrees of freedom are dominant.

Note that in most low energy nuclear
physics applications of the meson-exchange picture absorptive
effects are included, eiher via explicit treatment or by means of
simply neglecting the $\delta({\bf r})$-contribution in the respective
pion-exchange potentials. Also, these absorption effects should lead
to a modification of the $t$-dependence of the process $e + p
\rightarrow X + n$ for small $x$, namely the cross section would not
behave $\propto t$ at $t \rightarrow 0$ due to interference of the
pion exchange and the pion-Pomeron cut. This effect is well known in
strong interaction physics, for a review see Ref.~\cite{boresk}.

\section{CONCLUSIONS}

In this work, we have updated the analysis of the Sullivan mechanism,
i.e., the contribution of the nucleon's virtual meson cloud to the
deep inelastic structure functions, by using various recent improved
parton distribution functions fit to the host of inclusive, deep
inelastic lepton and neutrino scattering data. We have taken into
account contributions from the two lightest mesons, the $\pi$ and the
$K$, and, at various $Q^2$, we have separately adjusted the form
factors in the octet and decuplet channels and in the strange and
non-strange sectors to the $SU(3)$ and $SU(2)$ flavor breaking
components of the nucleon's antiquark sea and, simultaneously,
to its strange quark content.

We find that we can only achieve a good description of the $SU(3)$ as
well as the $SU(2)$ flavor asymmetry of the nucleon's antiquark sea
if we allow the $\pi NN$ and $\pi N\Delta$ vertices to differ
significantly, and that, in order to be able to also describe the
tail of the
strange quark component, we have to use harder cut-offs in the kaonic
sector. While the agreement of our calculations
with the data-based parametrizations is satisfactory and scale
independent for the flavor breaking share of the nucleon's
antiquark distributions, the corresponding flavor singlet component
is quite poorly described in the convolution picture.
This stresses the importance of gluonic degrees of freedom, even
at such a low scale as $Q^2=1$ GeV$^2$.

In particular, our analysis suggests exponential cut-off masses of
$\Lambda^e_{\pi NN} \approx 1000$ MeV, $\Lambda^e_{\pi N\Delta}
\approx 800$ MeV and $\Lambda^e_{KNY} \approx 1200$ MeV,
respectively. These results are based on two recent empirical parton
distribution functions, MRS(A') and CTEQ3M, which agree well with
both the determination of the nucleon's strange quark content
by the CCFR collaboration, related to the $SU(3)$ breaking, and the
measurement of the $SU(2)$ asymmetry by the NA51 collaboration.
Due to our inability to clearly distinguish between gluonic and
meson cloud contributions in the flavor singlet channel, the
$\Lambda^e_{\pi NN}$ and $\Lambda^e_{KNY}$ given here are upper
bounds of those quantities, while the $\Lambda^e_{\pi N\Delta}$ has
to be understood as a lower limit.

Our findings are in qualitative agreement with the faster fall-off
of the $p \to \Delta$ electromagnetic transition form factor as
compared to the proton's electromagnetic form factor, and they
suggest a sound breaking of the na\"\i ve $SU(6)$ symmetry relating
the quark-substructure of the nucleon and the $\Delta$-isobar.
Also, the vertices that we obtain from our analysis of the deep
inelastic scattering process are still significantly softer than
those employed in most effective $NN$ potentials fit to the rich
body of experimental phase shift data, although the discrepancy we
find is smaller than that quoted in early works in this context. They
correspond to a typical quark confinement size of about 0.6 fm.

Note, however, that the meson loop momenta that are probed in the
deep inelastic process investigated here, $|{\bf k}| \approx 0.8$
GeV, are very different from those relevant for low-energy nuclear
physics phenomena, as, for instance, in the meson-exchange
descriptions of the $NN$ interaction, and the respective distances
are smaller than the typical confinement size. This indicates the
limitations of applicability of the physical picture of a meson
cloud around the nucleon.

\acknowledgments{We wish to thank J.M.~Eisenberg and M.~Sargsyan for
many useful discussions.  This work was supported in part by the
Israel-USA Binational Science Foundation Grant No.~9200126, by the
MINERVA Foundation of the Federal Republic of Germany, and by the
U.S.~Department of Energy under Contract No.~DE-FG02-93ER40771.}


\end{document}